\PassOptionsToPackage{dvipsnames}{xcolor}
\documentclass[sigplan, authorversion, screen]{acmart}

\usepackage[all]{nowidow}
\usepackage{xcolor}
\usepackage{subcaption}
\usepackage{hyperref}
\usepackage{acronym}
\usepackage{listings}

\lstdefinelanguage{YAML}{ %
  keywords={true, false, null, on, off},
  keywordstyle=\color{blue},
  identifierstyle=\color{black},
  sensitive=false,
  comment=[l]{\#},
  commentstyle=\color{gray}\itshape,
  stringstyle=\color{purple},
  morestring=[b]',
  morestring=[b]"
}

\usepackage[capitalise,noabbrev]{cleveref}
\crefformat{section}{\S#2#1#3} %
\crefformat{subsection}{\S#2#1#3}
\crefformat{subsubsection}{\S#2#1#3}
\crefrangeformat{section}{\S#3#1#4 to~\S#5#2#6}
\crefrangeformat{subsection}{\S#3#1#4 to~\S#5#2#6}
\crefrangeformat{subsubsection}{\S#3#1#4 to~\S#5#2#6}

\usepackage[disable]{todonotes}

\begin{document}

\author{Mohammadreza Malekabbasi}
\affiliation{%
    \institution{TU Berlin}
    \city{Berlin}
    \country{Germany}}
\email{mm@3s.tu-berlin.de}
\orcid{0009-0002-0627-3600}

\author{Minghe Wang}
\affiliation{%
    \institution{TU Berlin}
    \city{Berlin}
    \country{Germany}}
\email{mw@3s.tu-berlin.de}
\orcid{0009-0001-3780-5828}

\author{David Bermbach}
\affiliation{%
    \institution{TU Berlin}
    \city{Berlin}
    \country{Germany}}
\email{db@3s.tu-berlin.de}
\orcid{0000-0002-7524-3256}

\title{DisCEdge: Distributed Context Management for Large Language Models at the Edge}


\acmYear{2026}\copyrightyear{2026}
\setcopyright{cc}
\setcctype[4.0]{by-nc-nd}
\acmConference[EuroMLSys '26]{The 6th Workshop on Machine Learning and Systems}{April 27--30, 2026}{Edinburgh, Scotland Uk}
\acmBooktitle{The 6th Workshop on Machine Learning and Systems (EuroMLSys '26), April 27--30, 2026, Edinburgh, Scotland Uk}
\acmDOI{10.1145/3805621.3807656}
\acmISBN{979-8-4007-2605-7/26/04}

\begin{abstract}
    Deploying Large Language Model (LLM) services at the edge benefits latency-sensitive and privacy-aware applications.
    However, the stateless nature of LLMs makes managing user context (e.g., sessions, preferences) across geo-distributed edge nodes challenging.
    Existing solutions, such as client-side context storage, introduce network latency and bandwidth overhead, undermining edge deployment advantages.

    We propose \textit{DisCEdge}, a distributed context management system that stores and replicates user context in tokenized form across edge nodes.
    By maintaining context as token sequences, our system avoids redundant computation and enables efficient data replication.
    We evaluate an open-source prototype in a realistic edge environment.
    DisCEdge improves median response times by up to 14.46\% and lowers median inter-node synchronization overhead by up to 15\% compared to a raw-text-based system.
    It also reduces client request sizes by a median of 90\% compared to client-side context management, while guaranteeing data consistency.
\end{abstract}

\maketitle

\section{Introduction}
\label{sec:introduction}
Large language models (LLMs) are becoming key enablers for intelligent, context-aware mobile applications.
Their ability to understand and leverage context differentiates them from simpler AI models.
However, their effectiveness critically depends on managing this context, a primary challenge in distributed edge environments.
While edge AI has shown promising results in using LLMs for robotics~\cite{black2410pi0, chen2025chatfly}, significant challenges remain in extending these capabilities to mobile and autonomous systems such as life-saving drones~\cite{searchwing}, autonomous vehicles~\cite{qu2025mobile}, and smartphones~\cite{yin2024llm}.
These challenges stem from a fundamental trade-off: commodity hardware, reliable latency, and a singleton deployment cannot all be achieved simultaneously~\cite{chen2024fogros2-plr}.
Thus, relying on a single LLM instance—whether on-device, at the edge, or in the cloud—often fails to meet the strict demands of mobile and resource-constrained environments~\cite{qu2025mobile}.

The trend of offering AI inference as a service follows serverless computing paradigms, where different models can be dynamically loaded and shared among multiple clients.
The push to migrate LLM inference to the edge is driven by the need for low latency and privacy.
However, edge devices have limited resources, hindering the performance of large models, especially when multiple models must run concurrently to serve complex applications~\cite{qu2025mobile, ichnowski2023fogros2}.
While many studies have explored cloud-and-device~\cite{xu2024edgellm, schafhalter2025bandwidth, ichnowski2023fogros2, narayan2025minions, wang2024cloud,mendula2024furcifer} and edge~\cite{zhang2024edgeshard} collaboration, the practical deployment of fully distributed LLM inference services at the edge remains largely conceptual~\cite{qu2025mobile}.

A fundamental barrier to such distributed deployments is context management.
LLMs are stateless by design; context—such as user sessions, preferences, and regional data—must be managed externally.
In a geo-distributed system, this context must be replicated across nodes to ensure a consistent user experience, which is notoriously difficult as it requires balancing high availability with strong data consistency across unreliable networks.
Managing context across a distributed fleet of edge nodes presents a far greater challenge than for single-node LLM services~\cite{yin2024llm}.

To the best of our knowledge, existing approaches do not adequately ensure that user interactions maintain a consistent context across geo-distributed edge nodes.
A common alternative, client-side context storage, not only places the burden of context management on the developer—often considered a clumsy approach~\cite{dong2024creating}—but also introduces additional network latency and increases bandwidth consumption, as the context must be sent with every request.
This is particularly problematic for mobile clients, where network constraints are a major limitation~\cite{schafhalter2023leveraging, ichnowski2023fogros2}, negating the latency benefits of edge deployment.
Conversely, inconsistent or stale context can lead to fragmented sessions and irrelevant LLM responses, undermining application reliability.
Addressing this gap requires a novel approach for managing context in geo-distributed LLM inference services.

We propose \emph{DisCEdge}, a distributed context management system for LLM inference services at the edge.
Our system replicates context as tokenized sequences to avoid redundant text-to-token processing during inference and reduce user data exposure.
Our experiments show that this approach effectively manages user context across geo-distributed LLM nodes, reducing median response times by 14.46\%, synchronization network overhead by 15\%, and client-to-edge network usage by 90\% compared to baseline approaches.
\footnote{We use the term LLM throughout this paper for simplicity, though our approach applies to other context-aware foundation models that use tokenization.}

Our contributions are as follows:
\begin{itemize}
    \item We design a distributed context management system for LLM inference services at the edge, enabling efficient replication of user context across geo-distributed nodes (\cref{sec:system_design}).
    \item We implement and open-source a proof-of-concept prototype and the corresponding evaluation setup (\cref{sec:prototype})\footnote{\href{https://github.com/ChaosRez/llm-context-management}{https://github.com/ChaosRez/llm-context-management}}.
    \item We evaluate the system in a realistic edge environment on commodity hardware, demonstrating reduced response times and network usage compared to edge-side raw text and client-side context storage approaches (\cref{sec:experiments}).
\end{itemize}

\section{Background and Related Work}
\label{sec:background}
To address context management for LLMs at the edge, we review existing strategies (\cref{sec:llms}) and geo-distributed storage systems (\cref{sec:storage}).

\subsection{Context Management in LLMs}
\label{sec:llms}
LLMs are inherently stateless, requiring all relevant context (e.g., session history) to be explicitly provided with each inference request.
While instruct models often handle single-turn tasks, chat models rely on maintaining multi-turn conversational flow (system, user, and assistant roles).
This session continuity is critical for applications like virtual assistants, autonomous driving, and function calling~\cite{kavathekar2025small}.
Ideally, this continuity should be handled transparently by the system and hidden behind the API~\cite{dong2024creating}.

\subsubsection{Constraints and Opportunities}
A primary system constraint is the model's limited \textit{context window}—ranging from 2K tokens in smaller models~\cite{zhang2024tinyllama} to 1M+ in large proprietary ones~\cite{comanici2025gemini}.
Inputs exceeding this limit must be truncated or summarized~\cite{wang2025recursively, mutasodirin2021investigating}.
This area of research is actively evolving, with ongoing efforts to develop models with larger context windows~\cite{martineau2024context}, extend existing ones~\cite{ding2024longrope}, and develop techniques to reduce context size while preserving semantic meaning~\cite{fei2023extending, extendContextWindow2024niederfahrenhorst}.

Crucially, text must be \textit{tokenized} into numeric sequences before processing.
Tokenization is model-dependent and compute-intensive, particularly for large texts or multi-modal inputs~\cite{wen2025token}.
However, managing context in tokenized form offers significant systems advantages: tokens are more compact than raw text and can be concatenated without re-processing.
Unlike embeddings used in Retrieval Augmented Generation (RAG)~\cite{gao2023retrieval}, tokens are the direct input format for inference, allowing for efficient pre-processing and caching.

\subsubsection{Context at the Edge}
Existing work on edge LLMs focuses on single-node inference efficiency via quantization, pruning, and memory optimization~\cite{chai2025flexquant, xu2024edgellm, liu2025adaptive, zheng2025review, ye2025jupiter}, or local context switching~\cite{yin2024llm, yu2025stateful}.
However, the challenge of maintaining consistent user context across \textit{distributed} edge nodes for roaming users remains largely unaddressed.

\subsection{Geo-Distributed Storage Systems}
\label{sec:storage}
Edge storage systems address the CAP theorem trade-off~\cite{brewer2000towards} under network constraints.
Systems like FogStore~\cite{gupta2018fogstore} offer ``differential consistency'' based on geographical relevance.
More recent systems like FReD~\cite{pfandzelter2023fred} provide client-centric consistency, where clients declaratively specify replication schemes that FReD then executes.
A client-side middleware intercepts requests to ensure consistency guarantees—such as monotonic reads and read-your-writes for mobile clients—originally proposed by Bermbach et al.~\cite{bermbach2013middleware}.
FReD nodes exchange data directly via peer-to-peer communication, using a naming service only for metadata and configuration.
Clients can dynamically choose which storage node to connect to and move between storage nodes as they roam.
FReD groups keys into \textit{keygroups} for which replication and consistency settings can be independently configured.

\subsubsection{Limitations for LLM Context}
These general-purpose storage solutions are not ideal for LLM context management for three reasons.
First, their client-centric consistency models are designed for scenarios where the client itself is mobile and directly interacts with the storage layer.
In our architecture, however, the data owner (the mobile user) and the entity managing storage (the static context manager on the edge node) are distinct.
Second, delegating replication complexity to the client contradicts our goal of a transparent, centralized-like service interface.
Third, existing systems do not exploit the specific properties of LLM context: it is sequential, monotonically growing, and beneficial to store in pre-tokenized form.
Our work proposes a specialized system that leverages these characteristics to provide transparent consistency for mobile users on edge infrastructure.

\section{DisCEdge Architecture}
\label{sec:system_design}
\begin{figure}[ht]
    \centering
    \includegraphics[width=\linewidth]{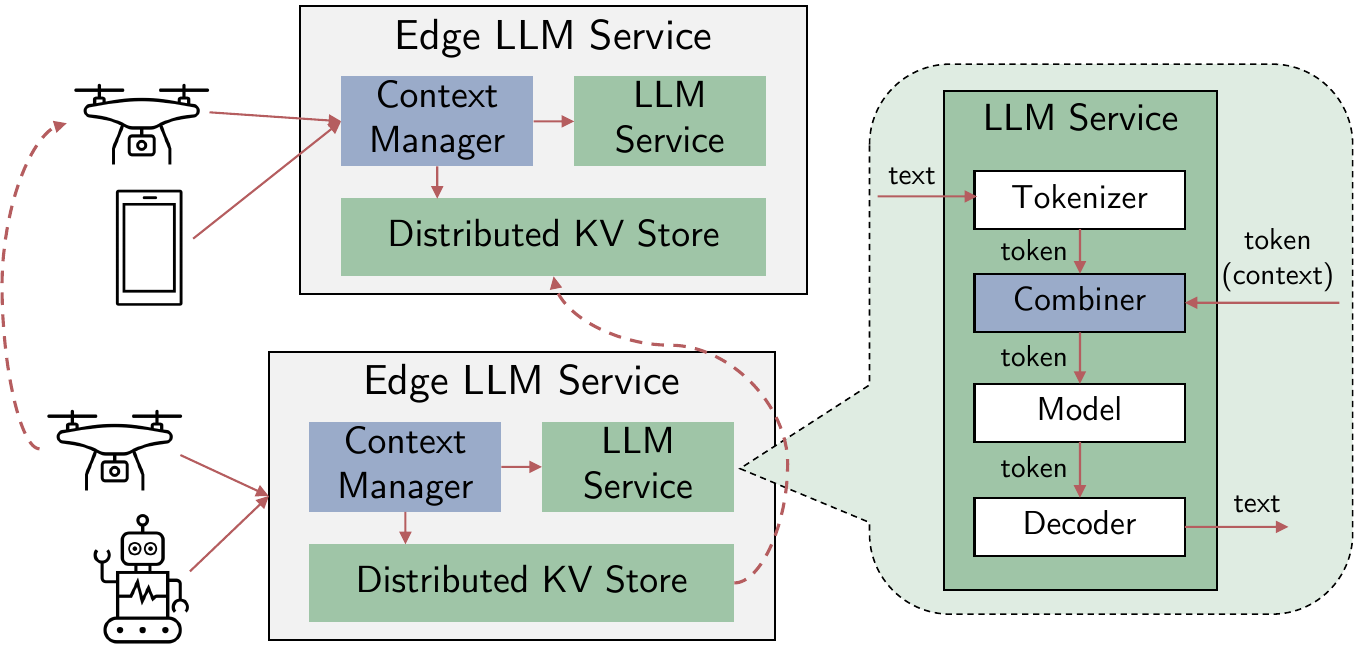}
    \caption{DisCEdge Architecture Overview. The Context Manager handles context retrieval and updates, while the LLM Service processes tokenized context directly.}
    \label{fig:system_architecture}
    \label{fig:llm_architecture}
\end{figure}

As shown in \cref{fig:system_architecture}, DisCEdge consists of multiple edge nodes, each containing a \textit{Context Manager} (\cref{sec:context_manager}), an \textit{LLM Service} (\cref{sec:llm_service}), and a \textit{Distributed KV Store} (\cref{sec:kv_store}), accessed by \textit{LLM clients} (\cref{sec:llm_clients}).

DisCEdge, while decentralized, provides a cloud-like interface without client-side replication complexity.
It employs lightweight tokenized context storage, which is more compact than raw text, reducing inter-node network overhead.
Tokens can be concatenated without repeated tokenization, avoiding redundant text-to-token processing for session history.
This optimization occurs earlier in the inference pipeline (before the \textit{prefill} phase), making it complementary to embedding-based techniques like RAG.

We focus on managing dynamic user session context, whose consistency is a significant challenge in geo-distributed systems.
Other context types, such as static preferences or regional data, have simpler consistency requirements and are easily extendable.

\subsection{Context Manager}
\label{sec:context_manager}
The Context Manager acts as an intelligent middleware between the client and the LLM Service, managing the lifecycle of user context.

Upon receiving a request, it assigns user and session identifiers if absent.
To provide strong consistency on top of the KV store's eventual consistency, DisCEdge leverages a lightweight, client-driven protocol.
The Context Manager uses a client-provided turn counter to verify its local session context is up-to-date before forwarding the request.
This approach is well-suited for mobile scenarios, as the client is the ultimate source of truth for the interaction sequence, ensuring session integrity as users roam between edge nodes.

By maintaining session context in pre-tokenized form, the Context Manager rapidly constructs context-aware prompts, eliminating repeated tokenization overhead.

\subsection{LLM Service}
\label{sec:llm_service}
The LLM Service executes language models, receiving a pre-tokenized context and user prompt from the Context Manager.
It is runtime and hardware agnostic, requiring only the ability to process token sequences and serve the same models (and tokenizers) as other nodes.

The LLM Service directly handles tokenized data, merging it with the user prompt (\cref{fig:llm_architecture}).
By only tokenizing the smaller, new input and concatenating it with the pre-tokenized context, the system avoids redundant processing, accelerating responses.

\subsection{Distributed KV Store}
\label{sec:kv_store}
The persistence layer uses a geo-distributed key-value (KV) store designed for edge environments, such as FReD~\cite{pfandzelter2023fred} or FogStore~\cite{gupta2018fogstore}.
It manages user context across edge nodes, distinct from the LLM's internal KV-cache.

When a session context is updated, the Context Manager writes to its local KV replica, which replicates the data to nodes serving the same model.
Each context has a time-to-live (TTL) for automatic cleanup.
Upon a request to a new node, the Context Manager reads the local replica; if stale (based on the turn counter), it retries, waiting for replication.

While many edge stores rely on client-side middleware to ensure consistency for mobile clients~\cite{bermbach2013middleware}, DisCEdge delegates this to the Context Manager for transparency.
The consistency-availability trade-off is client-configurable: applications demanding strong consistency can receive failure notifications if synchronization is prevented, while those prioritizing availability can proceed with potentially stale context.

\subsection{LLM Clients}
\label{sec:llm_clients}
Clients use the same request format as the standard LLM Service, adding user and session identifiers.
To enable our lightweight consistency protocol, the client maintains and sends a simple turn counter with each request within a session.
We assume clients can determine the closest edge node using either a service registry or geo-aware routing approach introduced in GeoFaaS~\cite{malekabbasi2024geofaas}.

\section{Evaluation}
\label{sec:evaluation}
To showcase DisCEdge, we implement a proof-of-concept prototype and conduct experiments evaluating its performance in an edge environment.

\subsection{Prototype Implementation}
\label{sec:prototype}

We implement DisCEdge in Go atop LLaMA.Cpp~\cite{gerganov2025llamacpp}, modifying it to accept pre-tokenized context via the \texttt{/completion} API\footnote{\href{https://github.com/ChaosRez/llama.cpp-fastencode}{https://github.com/ChaosRez/llama.cpp-fastencode}}.
This optimization avoids re-tokenizing history, allowing the engine to only tokenize new prompts.
Updates to the context occur asynchronously after response generation.
The system supports three modes: (i) \textbf{raw} text, (ii) \textbf{tokenized}, and (iii) \textbf{client-side} (no edge storage, resembling standard API usage).

We opted for a self-hosted solution as commercial clouds~\cite{cloudflare_workers_ai, replicate, vertex_ai, hugging_face_endpoints}, to our knowledge, lack low-level tokenization access.
Context is stored in FReD, a distributed in-memory key-value store, with data isolated per model.
The Context Manager ensures consistency by validating retrieved context versions against local state (turn) and retrying if the data is outdated.

\subsection{Experiments}
\label{sec:experiments}
We evaluate DisCEdge through two main experiments.
The first validates our core design choice of using a tokenized context representation by quantifying its impact on latency, throughput, and network overhead against a raw text approach.
The second showcases the superiority of our edge-side architecture for mobile clients by comparing its end-to-end performance and network efficiency against client-side context management.

The context maintained in our experiments only contains user sessions, which is a sequence of chat \textit{turns}, as managing other context types is easily extensible.
We repeat experiments three times and report results with a 95\% confidence interval.
We use a candid sample 9-turn prompt scenario with dependence on previous turns to verify relevance of outputs, focusing on the performance of the context management system.
For consistency guarantee settings, we set the retry count to 3, each with a 10ms back off, though through all our experiments the Context Manager never needs to retry more than two times.
To measure edge node synchronization network overhead, we use \texttt{tcpdump} to capture network traffic on the specific port used by FReD nodes for peer communication and \texttt{tshark} to analyze packets.
Although it captures some additional packets, such as TCP handshakes, we decide not to perform intrusive network monitoring by modifying the FReD codebase.

The experimental setup consists of two edge nodes (an Nvidia Jetson TX2 and a Mac M2) and one client device (a Raspberry Pi 4), linked via a local network with negligible latency ($\le 1$ms).
To mimic geographically distributed edge infrastructure and observe synchronization overhead, we use the macOS Packet Filter (\texttt{pf}) tool to inject a 10ms one-way latency (20ms RTT) between edge nodes.
We opt for this controlled latency simulation over a physical WAN deployment to eliminate asymmetric network conditions during client handovers.
We observe significantly lower response times on the M2 node compared to the older TX2 node, as LLaMa.Cpp is optimized for Apple Silicon.
This setup simulates a typical edge computing environment for latency-sensitive applications.
We use 4-bit quantization \texttt{Q4\_K\_M} of Qwen1.5-0.5B-Chat model as a balance between performance and compactness.
We set the seed to \texttt{123}, temperature to \texttt{0}, a max token generation of \texttt{128}, and verify the number of generated tokens for all experiments to ensure consistent results.

\subsubsection{Edge-side Context Management: Tokenized  vs. Raw Text}
\label{sec:tokenized_vs_raw}
This experiment evaluates storing context as raw text versus pre-tokenized data to quantify the performance gains of avoiding repeated tokenization.
We compare two modes: (i) \textbf{raw} text, where the edge stores the conversation history as plain text and tokenizes the entire context with each new prompt, and (ii) \textbf{tokenized}, where the edge stores the context in its tokenized form.
We run a client session against a single edge node (once on M2 and once on TX2) for each mode.
We measure \textbf{End-to-End Response Time} to assess user-perceived latency (a primary goal for edge-based systems) and \textbf{Synchronization Network Overhead} to demonstrate the benefit of a compact context representation on inter-node traffic (crucial for scalability in edge environments).

\subsubsection*{Result}
\begin{figure*}[t]
    \centering
    \begin{subfigure}[t]{0.55\textwidth}
        \centering
        \includegraphics[width=\linewidth]{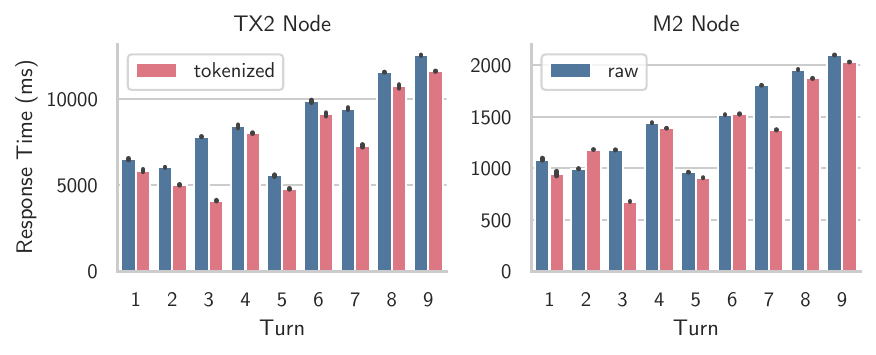}
        \caption{Client-observable response time per turn (error bars show 95\% CI).}
        \label{fig:tokenized_vs_raw_latency}
    \end{subfigure}
    \hfill
    \begin{subfigure}[t]{0.43\textwidth}
        \centering
        \includegraphics[width=\linewidth]{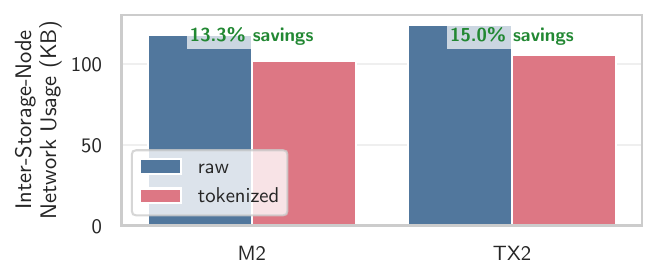}
        \caption{Inter-node synchronization overhead collected on M2.}
        \label{fig:tokenized_vs_raw_network}
    \end{subfigure}

    \caption{Tokenized vs.\ raw text edge-side context management performance on each edge node.}
    \label{fig:tokenized_vs_raw}
\end{figure*}

As shown in Figure~\ref{fig:tokenized_vs_raw_latency}, tokenized context storage outperforms raw text storage in median response times (14.46\% speedup on TX2, and 8.75\% on M2).
Note that, since the prompt in each turn is different, and thus can have a different length and complexity, the response times are not perfectly linear with the number of tokens in the context.
However, the overall trend shows that tokenized context storage consistently leads to lower response times compared to raw text storage.
This is because tokenization reduces the overhead of processing large text inputs.
We also observe that the number of tokens processed per second (\textbf{TPS}) is slightly higher for tokenized context storage compared to raw text storage (2.85\% speedup on TX2, 1.41\% on M2), but also decreases with the growth of context.
While the TX2 is still much slower than the M2 node, tokenized requests are significantly faster than client-side ones, especially when the context is much larger than the prompt.
While the asynchronous tokenization step by the Context Manager takes 4ms to 50ms on the TX2 node and is consistently <1ms on the M2 node, this step is performed asynchronously with sending the response to the client.
This decreases the client-observable response times.

According to \cref{fig:tokenized_vs_raw_network}, the tokenized context storage reduces the network usage by 13.3\% and 15\% on M2 and TX2 nodes, respectively.
This is because tokenized context storage reduces the size of the context data that needs to be synchronized between edge nodes, leading to lower network overhead.

\subsubsection{Edge-Side vs. Client-Side Context Management for Mobile Clients}
\label{sec:moving_clients}
This experiment evaluates DisCEdge in a mobile client scenario, a key use case for edge computing.
We compare our proposed edge-side approach (using tokenized context) against a baseline using client-side context management, where the client sends the complete context with each request.
As mentioned in \cref{sec:introduction}, This baseline is common but can be inefficient for mobile devices with limited bandwidth.
We simulate client mobility by having the client alternate between two different edge nodes after two turns during a conversation.
We measure \textbf{End-to-End Response Time} to show that DisCEdge, while maintaining consistency, is faster for the user than transmitting the full context from the client, and \textbf{Client-to-Edge Network Usage} to quantify the reduction in data sent from the client.

\subsubsection*{Result} %
\begin{figure*}[t]
    \centering
    \begin{subfigure}[t]{0.62\textwidth} %
        \centering
        \includegraphics[width=\linewidth]{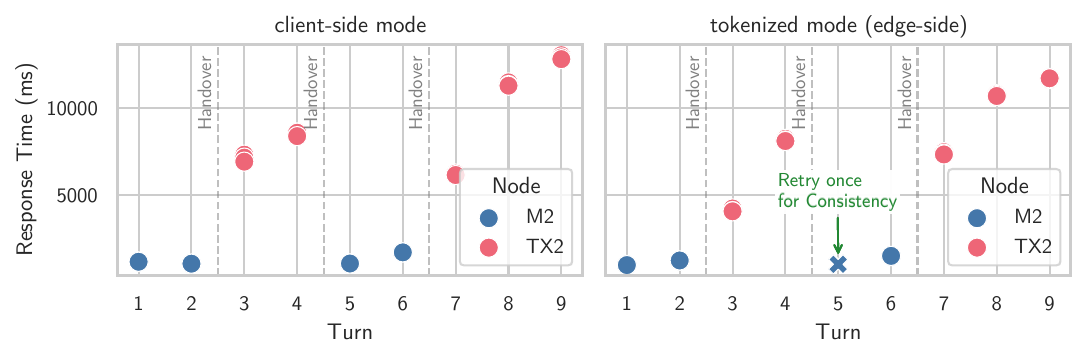}
        \caption{Response time per turn.}
        \label{fig:response_time_moving}
    \end{subfigure}
    \hfill
    \begin{subfigure}[t]{0.37\textwidth} %
        \centering
        \includegraphics[width=\linewidth]{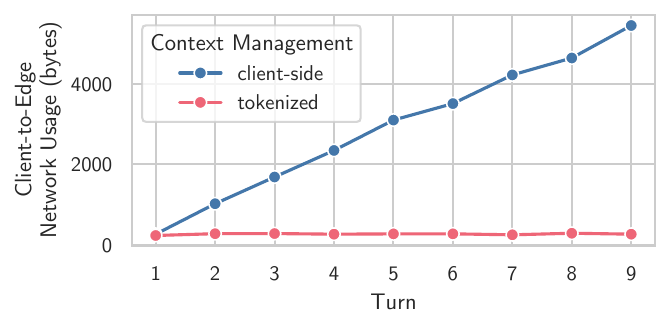}
        \caption{Request size per turn.}
        \label{fig:network_usage_moving}
    \end{subfigure}

    \caption{Mobile client performance: (a) DisCEdge outperforms client-side context despite handovers at turns 3, 5, and 7. (b) DisCEdge keeps request size constant (client-side grows linearly).}
    \label{fig:moving_client_combined}
\end{figure*}

As depicted in \cref{fig:response_time_moving}, DisCEdge reduces response time compared to the client-side baseline.
Our approach achieves a median speedup of 5.93\%, with a 2.51\% improvement on the M2 node and a 6.29\% improvement on the TX2 node.
The benefits in terms of network usage are even more substantial.
\cref{fig:network_usage_moving} shows that our system reduces the client request size by a median of 90\%.
This is because our edge-side approach maintains a constant request size (dependent only on the new prompt), whereas the client-side approach leads to a linear growth in request size with every turn as the entire conversation history is transmitted.
This drastic reduction in client-to-edge network traffic is a critical advantage for mobile clients, where bandwidth is often limited and costly.

We observe greater benefits in network usage and response time as the context grows larger.
This demonstrates that DisCEdge ensures context consistency for moving clients while significantly improving performance and minimizing network overhead.

\section{Discussion}
\label{sec:discussion}
Our evaluation demonstrates the benefits of edge-side, tokenized context management for LLM services at the edge.
However, our work has several limitations that present opportunities for future research.
Our experiments use a single client, which raises questions about scalability.
While managing each user's context as a separate key-value pair avoids storage contention, replication latency during client handovers remains a factor.
A comprehensive multi-tenant scalability analysis is an important next step.

Our evaluation compares against common baseline approaches—raw text and client-side storage.
We view more advanced techniques, such as context summarization, as complementary.
Summarization could prune context before storage in our tokenized format, helping manage very long-term histories.

A key threat to external validity is the use of a synthetic prompt scenario.
While designed to test growing context, real-world interactions are more varied, and prompt complexity influences response times.

Furthermore, our approach is subject to inherent LLM limitations, as models may struggle to use information buried in long contexts~\cite{wu2024reducing}.
Future work could explore selective retrieval or pruning strategies to mitigate this issue.

From an implementation perspective, DisCEdge is designed to complement existing single-node KV cache optimizers (e.g., vLLM~\cite{kwon2023efficient}, SGLang~\cite{zheng2024sglang}) rather than compete directly against them.
While migrating or sharing the internal KV cache of the LLM~\cite{gao2025rethinking, yu2025stateful} avoids re-processing tokens, it introduces a critical trade-off in distributed environments.
KV caches are massive, particularly for large models and long sessions, and replicating them across edge nodes would incur prohibitive network bandwidth overhead.
Therefore, DisCEdge purposely replicates the much smaller pre-tokenized context, making on-the-fly KV cache recomputation at new edge nodes preferable to transferring and storing large cache replicas.

Finally, our prototype does not implement an explicit eviction policy beyond TTL and client's signal.
For long-running applications, future work could explore cache eviction strategies, such as Least Recently Used (LRU) or session-based timeouts, to control storage overhead.

\section{Conclusion}
\label{sec:conclusion}
Deploying LLM services at the edge is beneficial for latency-sensitive and privacy-aware applications.
However, the stateless nature of LLMs makes managing user context across geo-distributed edge nodes challenging.
Existing solutions, such as client-side context storage, often introduce network latency and bandwidth overhead, undermining the advantages of edge deployment.

We proposed DisCEdge, a distributed context management system that stores and replicates user context in tokenized form across edge nodes.
By maintaining context as token sequences rather than raw text, our system avoided redundant computation and enabled efficient data replication.
We implemented and evaluated an open-source prototype in a realistic edge environment with commodity hardware.
We showed DisCEdge improved median response times by up to 14.46\% and lowered median inter-node synchronization overhead by up to 15\% compared to a raw-text-based storage.
It also reduced client request sizes by a median of 90\% compared to client-side context management, while guaranteeing data consistency.

\begin{acks}
    Funded by the \grantsponsor{BMFTR}{Bundesministerium für Forschung, Technologie und Raumfahrt (BMFTR, German Federal Ministry of Research, Technology and Space)}{https://www.bmftr.bund.de/EN/Home/home_node.html} -- \grantnum{BMFTR}{16KISK183}.
    We thank Tobias Pfandzelter for helping us with the FReD source code explanation, Mohammad Mohammadi for reviewing our changes to LLaMa.Cpp source code, and the volunteer reviewers for their valuable feedback.
\end{acks}

\balance

\bibliographystyle{ACM-Reference-Format}
\bibliography{bibliography.bib}

\onecolumn 
\appendix
\section{Experiment Details}
\label{sec:appendix_config}

This appendix provides the complete configuration details for the experiments described in \cref{sec:evaluation}.

\subsection{Prompt Scenario}
\label{sec:appendix_prompt}
The following YAML configuration shows the 9-turn prompt scenario used in our experiments.
The scenario simulates a technical conversation about robotics and autonomous systems, with questions that build upon previous responses to test context dependency.

\begin{lstlisting}[language=YAML, caption={9-turn prompt scenario for robotics and autonomous systems}, label={lst:config}]
name: "Robotics and Autonomous Systems Test"
model_name: "Qwen/Qwen1.5-0.5B-Chat"
user_id: "robotics_dev"
messages:
  1. "What are the fundamental components of an autonomous mobile robot?"
  2. "You mentioned sensors. What are the most common types for obstacle avoidance?"
  3. "Can you explain the concept of a PID controller in the context of motor control?"
  4. "Write a simple Python function for a proportional (P) controller."
  5. "In your previous code, what do the `kp` and `error` variables represent?"
  6. "How would you modify that function to include the integral (I) component?"
  7. "Now, let's talk about localization. What is SLAM?"
  8. "What are some of the main challenges when implementing that on a small, low-power robot?"
  9. "Can you compare the EKF SLAM and Particle Filter SLAM approaches?"
\end{lstlisting}

\subsection{Hardware Specifications}
\label{sec:appendix_hardware}
Table~\ref{tab:hardware} summarizes the hardware specifications of the devices used in our experiments.

\begin{table}[h]
\centering
\caption{Hardware specifications of experimental devices}
\label{tab:hardware}
\begin{tabular}{lll}
\hline
\textbf{Device} & \textbf{Specification} & \textbf{Role} \\
\hline
Nvidia Jetson TX2 & ARM Cortex-A57 (4-core) & Edge Node \\
                  & 8GB unified memory & \\
                  & 256-core Pascal GPU & \\
\hline
Apple Mac M2 & 8-core CPU (4P+4E) & Edge Node \\
             & 16GB unified memory & \\
             & 8-core GPU & \\
\hline
Raspberry Pi 4 & ARM Cortex-A72 (4-core) & Client Device \\
               & 4GB RAM & \\
\hline
\end{tabular}
\end{table}

\end{document}